# Demonstration of an electric field conjugation algorithm for improved starlight rejection through a single mode optical fiber

Jorge Llop Sayson
Garreth Ruane
Dimitri Mawet
Nemanja Jovanovic
Benjamin Calvin
Nicolas Levraud
Milan Roberson
Jacques-Robert Delorme
Daniel Echeverri
Nikita Klimovich
Yeyuan Xin

# Demonstration of an electric field conjugation algorithm for improved starlight rejection through a single mode optical fiber

Jorge Llop Sayson,[a,*] Garreth Ruane,[a] Dimitri Mawet,[a,b] Nemanja Jovanovic,[a] Benjamin Calvin,[a] Nicolas Levraud,[a,c] Milan Roberson,[a] Jacques-Robert Delorme,[a] Daniel Echeverri,[a] Nikita Klimovich,[a] and Yeyuan Xin[a]
[a]California Institute of Technology, Pasadena, California, United States
[b]Jet Propulsion Laboratory, California Institute of Technology, Pasadena, California, United States
[c]Institut d'Optique Graduate School, Palaiseau, France

**Abstract.** Linking a coronagraph instrument to a spectrograph via a single-mode optical fiber is a pathway toward detailed characterization of exoplanet atmospheres with current and future ground- and space-based telescopes. However, given the extreme brightness ratio and small angular separation between planets and their host stars, the planet signal-to-noise ratio will likely be limited by the unwanted coupling of starlight into the fiber. To address this issue, we utilize a wavefront control loop and a deformable mirror to systematically reject starlight from the fiber by measuring what is transmitted through the fiber. The wavefront control algorithm is based on the formalism of electric field conjugation (EFC), which in our case accounts for the spatial mode selectivity of the fiber. This is achieved by using a control output that is the overlap integral of the electric field with the fundamental mode of a single-mode fiber. This quantity can be estimated by pairwise image plane probes injected using a deformable mirror. We present simulation and laboratory results that demonstrate our approach offers a significant improvement in starlight suppression through the fiber relative to a conventional EFC controller. With our experimental setup, which provides an initial normalized intensity of $3 \times 10^{-4}$ in the fiber at an angular separation of $4\lambda/D$, we obtain a final normalized intensity of $3 \times 10^{-6}$ in monochromatic light at $\lambda = 635$ nm through the fiber (100× suppression factor) and $2 \times 10^{-5}$ in $\Delta\lambda/\lambda = 8\%$ broadband light about $\lambda = 625$ nm (10× suppression factor). The fiber-based approach improves the sensitivity of spectral measurements at high contrast and may serve as an integral part of future space-based exoplanet imaging missions as well as ground-based instruments. © *The Authors. Published by SPIE under a Creative Commons Attribution 4.0 Unported License. Distribution or reproduction of this work in whole or in part requires full attribution of the original publication, including its DOI.* [DOI: 10.1117/1.JATIS.5.1.019004]

## 1 Introduction

Directly detecting the spectral signatures of molecules in the atmosphere of exoplanets, including biosignatures on temperate Earth-size planets, poses an immense technical challenge. Noise due to stray starlight diffracted from the telescope aperture as well as static and dynamic aberrations throughout the optical system limits the detection significance of the planet's spectral features. Furthermore, the wavefront quality and stability requirements for detecting and characterizing Earth-size exoplanets around solar-type stars with space-based missions, such as the Habitable Exoplanet Observatory (HabEx)[1] and Large UV/Optical/IR Surveyor (LUVOIR)[2] mission concepts,[3] and around M-type stars with the next-generation giant segmented mirror telescopes on the ground, will be at the limits of current wavefront sensing and control techniques and technologies.[4]

Fiber-fed spectrographs have been used in astronomy since the 1980s.[5] In the last decades, advances in adaptive optics (AO) have enabled diffraction-limited imaging and spectroscopy with 8- to 10-m class ground-based telescopes and made the use of single-mode fibers (SMFs) an advantageous option.[6–8] Recently, we introduced a practical concept that allows for the spectroscopic characterization of known exoplanets by linking the final focal plane of a coronagraph to a spectrograph via a single-mode optical fiber.[9] A fiber injection unit (FIU) collects the known exoplanet's signal by coupling its light into an SMF. In most cases, the signal-to-noise ratio of the planet spectrum is limited by speckle and photon noise sources from starlight. Minimizing the stellar electric field that couples into the fiber reduces these noise sources such that the faint planet signal can be spectroscopically analyzed. The motivation for using an SMF is to exploit its mode selectivity to further reject unwanted starlight.

Wavefront control techniques aim to eliminate stellar speckles and reduce contamination of the companion's signal using AO. A deformable mirror (DM) placed at a pupil plane modifies the incoming wavefront to create a dark, speckle-free region in the image plane using one of several approaches that have been implemented successfully in previous laboratory demonstrations.[10] A notable example is the electric field conjugation (EFC) algorithm,[11] which is the baseline wavefront control algorithm for the WFIRST coronagraph instrument (CGI).[12] By finding the minimum of the electric field, EFC solves for the shape of the DM, which is characterized by $N \times N$ actuator heights.

Here, we introduce a new algorithm based on the EFC formalism that modifies the wavefront to minimize the speckles coupling into an SMF. We present the modified formalism of EFC that accounts for the modal selectivity of the SMF, results from simulations, as well as supporting laboratory experiments.

*Address all correspondence to Jorge Llop Sayson, E-mail: jllopsay@caltech.edu

## 2 Electric Field Sensing

EFC iteratively reduces stellar intensity in a region of the image plane using an estimate of the electric field. In the case of an SMF, the measured intensity at the output of the fiber is the overlap integral of the electric field at the input of the fiber multiplied by the fundamental mode of the fiber

$$I \propto \left| \int E_{\text{im}} \Psi_{\text{SMF}} da \right|^2, \quad (1)$$

where $E_{\text{im}}$ is the electric field, $\Psi_{\text{SMF}}$ is the fiber mode shape, and $da$ is the differential area element in the image plane. The control algorithm presented here relies on the sensing of the real and imaginary parts of the electric field through the mode of the fiber. The procedure for sensing the overlap integral is based on the pairwise probing method introduced by Give'On et al.[11] and further developed by Groff et al.[10] However, instead of sensing the field at a set of pixels, the resolution element in this case is the overlap integral for the SMF referred to in this work as a fibxel.

We write the electric field in the image plane as the output of the coronagraph operator $\mathcal{C}\{\cdot\}$

$$E_{\text{im}} = \mathcal{C}\{A \, e^{\alpha+i\beta} e^{i\phi}\}, \quad (2)$$

where $A$ is the pupil field, $\alpha$ and $\beta$ are the amplitude and phase aberrations, respectively, and $\phi$ is the phase delay introduced by the DM. Assuming small changes in DM shapes, we use a truncated Taylor series expansion about $\phi = 0$ to find the linear relationship between the DM actuator heights and the field at the fiber. That is,

$$E_{\text{im}} \approx \mathcal{C}\{A \, e^{\alpha+i\beta}\} + i\mathcal{C}\{\phi\} = E_{\text{Sp}} + Gu, \quad (3)$$

where $E_{\text{Sp}}$ is the speckle field we seek to sense, $G$ is the control matrix, or Jacobian of the system, and $u$ contains the changes in DM actuator heights. The intensity measured at the output of a fiber is

$$\begin{aligned} I &= \left| \int (E_{\text{Sp}} + Gu) \Psi_{\text{SMF}} da \right|^2, \\ &= \left| \int \Psi_{\text{SMF}} E_{\text{Sp}} da \right|^2 + \left| \int \Psi_{\text{SMF}} Gu \, da \right|^2 \\ &\quad + 2\text{Re}\left\{ \int \Psi_{\text{SMF}} E_{\text{Sp}} da \times \int \Psi_{\text{SMF}} Gu \, da \right\}. \end{aligned} \quad (4)$$

For a pair of probes, $\pm Gu$, the difference between intensities of the positive and negative probe images is

$$\begin{aligned} \Delta I &= 4\text{Re}\left\{ \int \Psi_{\text{SMF}} E_{\text{Sp}} da \times \int \Psi_{\text{SMF}} Gu \, da \right\}, \\ &= 4 \int \Psi_{\text{SMF}} \text{Re}\{E_{\text{Sp}}\} da \int \Psi_{\text{SMF}} \text{Re}\{Gu\} da, \\ &\quad + 4 \int \Psi_{\text{SMF}} \text{Im}\{E_{\text{Sp}}\} da \int \Psi_{\text{SMF}} \text{Im}\{Gu\} da. \end{aligned} \quad (5)$$

For $n$ different pairs of probes

$$\begin{bmatrix} \Delta I_1 \\ \vdots \\ \Delta I_n \end{bmatrix} = 4 \begin{bmatrix} \int \Psi_{\text{SMF}} \text{Re}\{Gu_1\} da & \int \Psi_{\text{SMF}} \text{Im}\{Gu_1\} da \\ \vdots & \vdots \\ \int \Psi_{\text{SMF}} \text{Re}\{Gu_n\} da & \int \Psi_{\text{SMF}} \text{Im}\{Gu_n\} da \end{bmatrix} \\ \times \begin{bmatrix} \int \Psi_{\text{SMF}} \text{Re}\{E_{\text{Sp}}\} da \\ \int \Psi_{\text{SMF}} \text{Im}\{E_{\text{Sp}}\} da \end{bmatrix}, \quad (6)$$

or more simply, $z = Hx$. Taking the pseudoinverse of the observation matrix $H$, we find an estimate of the fibxel electric field $\hat{x} = H^{-\dagger} z$, where $\hat{x}$ is specifically the estimate of the complex-valued overlap integral. This estimate is computed at each control iteration. For a system equipped with more than one SMF in the image plane, a larger number of fibxels is used in the matrices above.

## 3 EFC through a Single-Mode Fiber

Once the overlap integral of the electric field in the image plane is estimated, we use a similar approach to the conventional EFC algorithm.[11] Assuming a linear relationship between the DM actuators and field in the image plane [see Eq. (3)], we calculate the DM shape that minimizes, in the least squares sense with a cost function given by $W = |\int (E_{\text{Sp}} + Gu) \Psi_{\text{SMF}} da|^2$, the overlap integral. This is done by $u = -G^{-\dagger} \hat{x}$, where

$$\hat{x} = \begin{bmatrix} \int \Psi_{\text{SMF}} \text{Re}\{E_{\text{Sp}}\} da \\ \int \Psi_{\text{SMF}} \text{Im}\{E_{\text{Sp}}\} da \end{bmatrix}. \quad (7)$$

In conventional EFC, $G$ accounts for the effect of each actuator on the signal measured by pixels in the dark hole (DH). In the case of an SMF, $G$ accounts for the effect of each actuator on the overlap integral(s). Hence, similar to conventional EFC, $G$ is computed using a model of the optical system, where each DM actuator is poked and its effect on the overlap integral is stored in $G$. The computation of the shape of the DM, $u$, is done iteratively until the starlight coupling into the SMF is minimized. Here, we report the performance of the EFC algorithm in terms of normalized intensity.

## 4 Definitions of Normalized Intensity

For the sake of clarity, we define the following metrics used in this paper to evaluate contrast performance:

- **Mean normalized intensity.** The mean intensity in the DH divided by the peak intensity of the noncoronagraphic star PSF. This is a commonly used metric to measure and is often found in the literature as simply normalized intensity. In this paper, we will only use this definition in Sec. 7.1, in the context of conventional, camera-based EFC.
- **SMF normalized intensity.** The power measured at the output of the SMF divided by the intensity measured at the output of the SMF centered on the noncoronagraphic star PSF. This is the main metric we use in this paper to assess the performance of the new algorithm.

- **Pixel aperture normalized intensity.** The total intensity measured on an aperture on the camera of the size of the SMF, divided by the total intensity of the same aperture centered on the noncoronagraphic star PSF. We use this metric to effectively compare the new algorithm in terms of the intensity at the position of the fiber. This metric can be thought as the fiber normalized intensity of a multimode fiber, with an aperture the same size as the experiment's SMF.

The normalized intensity is equivalent to raw contrast when the throughput of the off-axis PSF at the angular separation of the planet is unaffected by the coronagraph.

## 5 Simulations

In order to validate the control algorithm presented above, we performed simulations in an end-to-end testbed simulation of the high contrast spectroscopy for segmented telescopes testbed (HCST)[13,14] in the Exoplanet Technology Laboratory (ET Lab) at Caltech where we have carried out the experimental tests described in Sec. 5.1. This model is based on a MATLAB code that uses the PROPER[15] library to perform realistic propagations for coronagraph and AO systems. This model assumes a point source for the star and static aberrations. We use surface errors of 3-nm RMS per optic with randomly generated error maps based on a power spectral density function, calculated from measurements of HCST's optics. The model for the SMF is a two-dimensional Gaussian of $1.4\lambda_0/D$ FWHM.

For monochromatic light at $\lambda = 650$ nm, we obtain almost perfect suppression of the coupling of the speckles through the SMF. In theory, the DM at the pupil plane has full control authority over the coupling of any monochromatic speckle through an SMF placed within the control area on the image plane.

We also simulated the new algorithm with polychromatic light with a $\Delta\lambda/\lambda = 10\%$ bandwidth, centered at $\lambda_0 = 650$ nm, and compared it to the performance of conventional EFC on the same setup and speckle field (see Fig. 1). In both simulations, we compute the power at the output of an SMF and the intensity read from the pixels of a simulated camera in the same image plane. We compare the SMF normalized intensity, calculated with the SMF, to the pixel aperture normalized intensity, calculated by integrating intensity over pixels (see Sec. 4). The performance of the new algorithm is consistently better in terms of final normalized intensity for different surface error maps on the optics than conventional EFC.

As we discussed in Sec. 3, the new algorithm does not try to eliminate the electric field at the fiber position; instead, it minimizes the overlap integral of the speckles with the fundamental mode of the SMF. Figure 2 shows the outcome of this important difference between the new algorithm presented here and conventional EFC. Although more light falls on the region of the SMF for the new algorithm [Fig. 2(b)] with respect to the conventional EFC result [Fig. 2(b)], there is less light coupling into the fundamental mode of the SMF, which is the ultimate goal. Since the cost function is less restrictive and the algorithm is not required to move the same amount of light from the region of the fiber, the DM strokes are also smaller. Two main factors cause the difference in DM stroke: (1) the modal selectivity of the SMF helps relieve the overall work of the DM and (2) conventional EFC uses several resolution elements to effectively suppress the diffracted starlight in the region of the fiber making its cost function more restrictive.

In Figs. 3 and 4, we zoom in on the region of the fiber and compare the intensity and phase, respectively, for both a conventional and fiber-based EFC example case. Figure 3 shows that conventional EFC tries to suppress the amplitude of the electric field, and thus the intensity, creating a DH in the stellar speckle field. On the other hand, the fiber-based algorithm leaves a

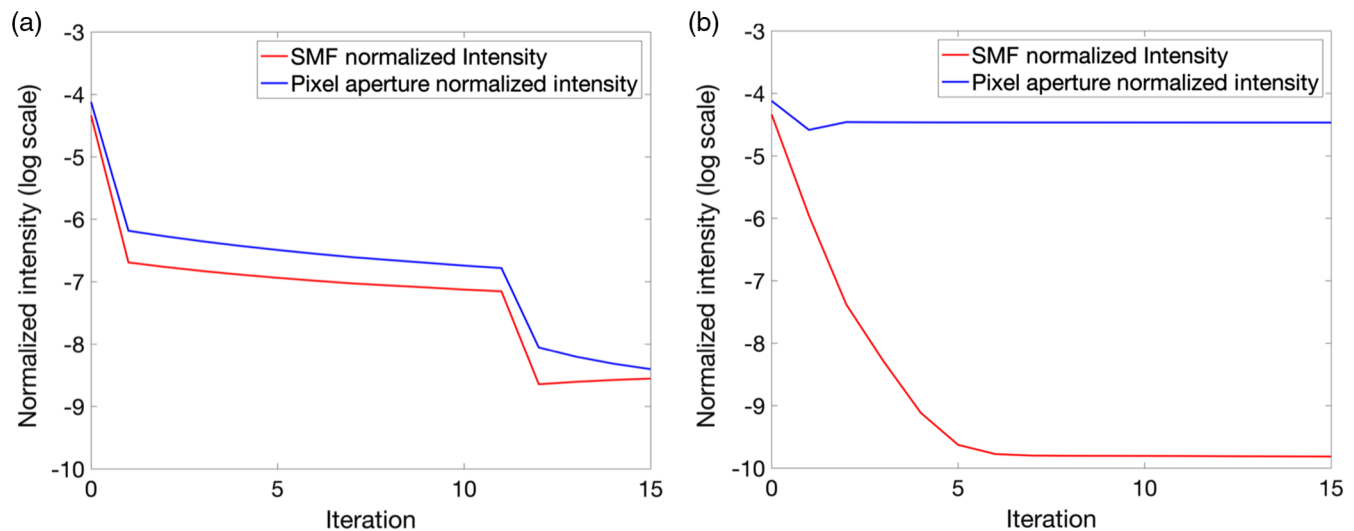

**Fig. 1** Simulations comparing (a) conventional EFC and (b) the new fiber-based algorithm. In both cases, the SMF normalized intensity (red line) is lower than the pixel aperture normalized intensity (blue line). The fiber-based EFC algorithm consistently yields deeper nulls in fewer iterations. The $G$ matrix is recomputed at iteration number 11 in both cases. In the conventional EFC case in (a), the improvement is clearly seen. For the new algorithm, in (b), there is no significant improvement after the recalculation of the $G$ matrix. All of the simulations assume polychromatic light with a spectral bandwidth of $\Delta\lambda/\lambda = 10\%$.

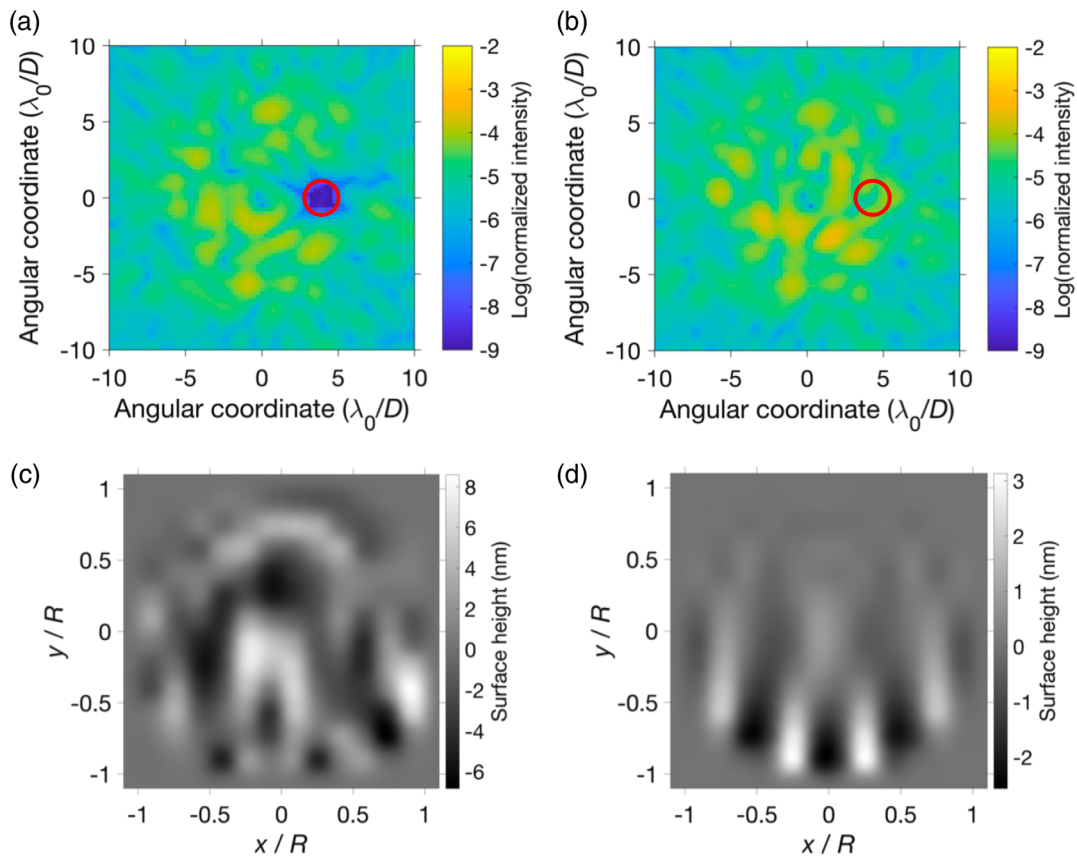

**Fig. 2** (a), (b) Simulated stellar PSFs in log normalized intensity after running (a) conventional and (b) fiber-based EFC. (c), (d) The corresponding DM shapes. The red circle indicates the control area for the case of conventional EFC and the position of the fiber for the case of fiber-based EFC, both centered at $4\lambda_0/D$ from the center of the PSF.

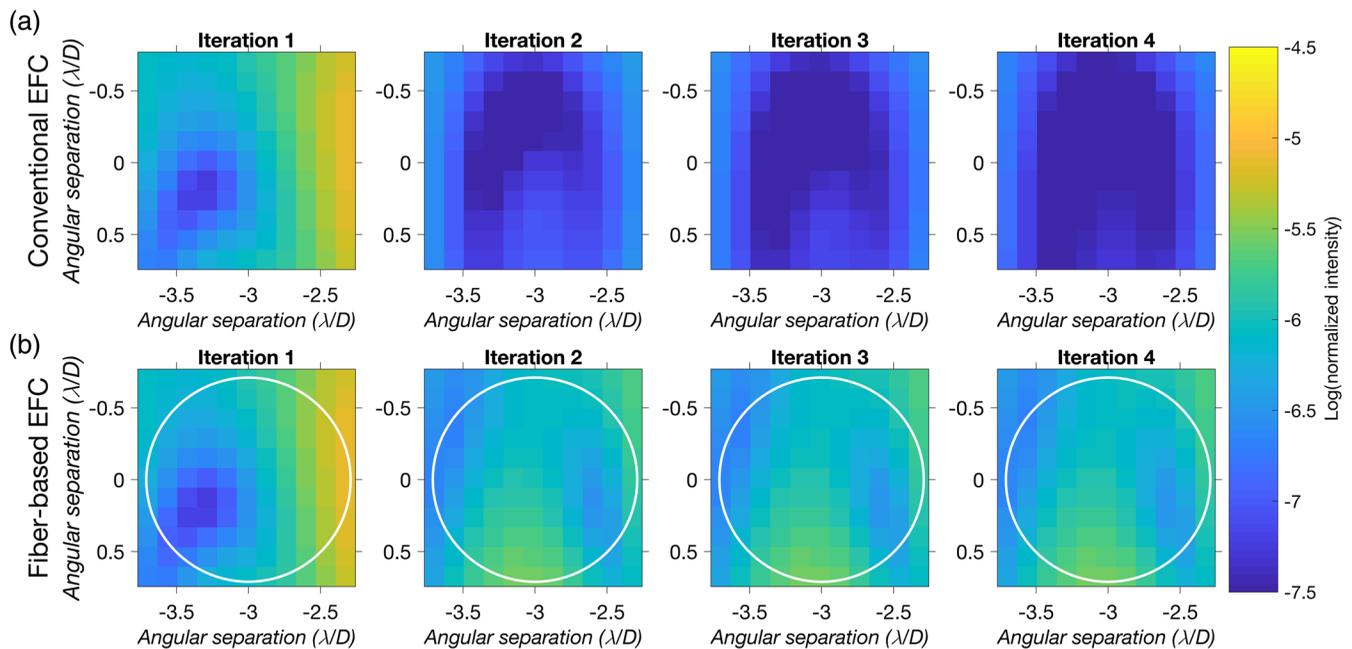

**Fig. 3** Zoom in on the region of the fiber for the simulations in Figs. 1 and 2. (a) The first four iterations on a conventional EFC run show how EFC tries to suppress the intensity over the region. However, (b) the new algorithm does not create a DH since it is only minimizing the overlap integral and converges to a solution is only a few steps. The white circle indicates the size of the mode of the fiber.

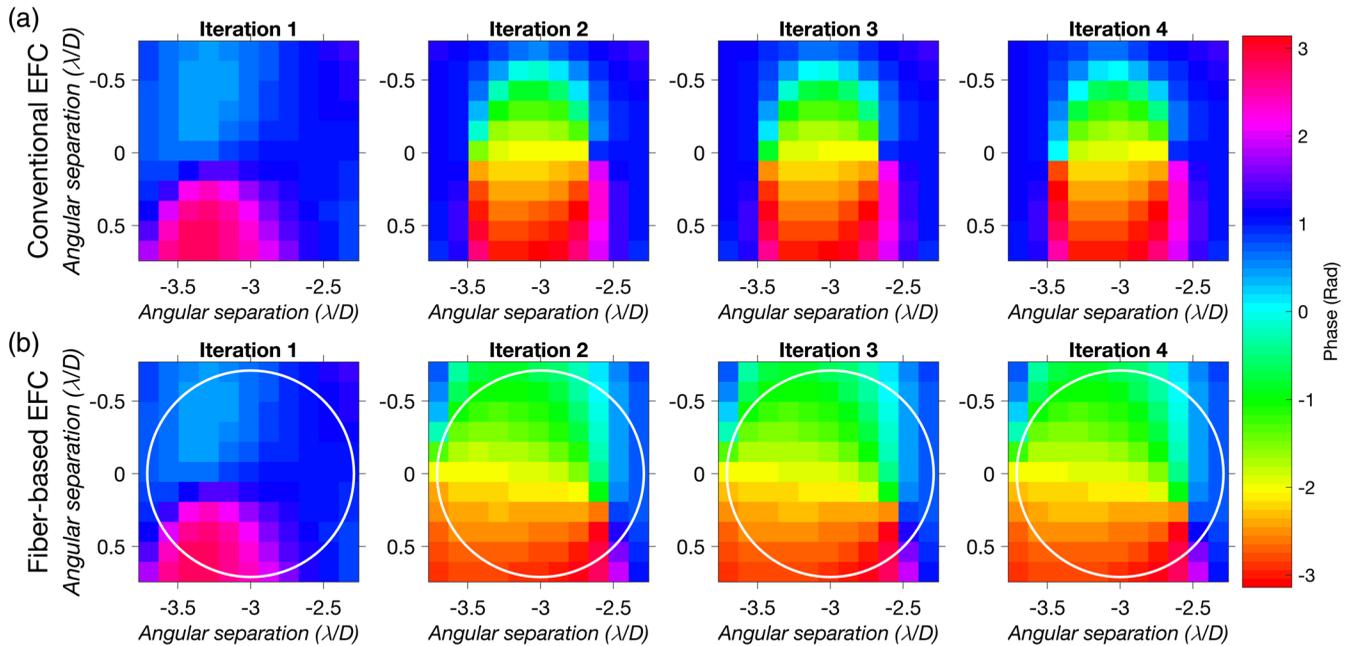

**Fig. 4** Same as Fig. 3, but showing the phase of the stellar field. (a) In the conventional EFC case, the phase over the region of the fiber does not obey any particular pattern since EFC works on suppressing the intensity at every position the region. On the other hand, (b) the fiber-based algorithm converges to a field that is asymmetric across the fiber tip.

small amount of stellar intensity at the fiber position. However, comparing the phase of the residual stellar fields (see Fig. 4) demonstrates that the fiber-based algorithm converges to a state where the phase becomes asymmetric or singular at the fiber tip preventing the starlight from coupling into the SMF.

### 5.1 Response to Tip-Tilt Errors

We analyze the sensitivity of the null to postcoronagraphic tip-tilt errors by adding offsets to the fiber before and after the nulling with the fiber-based EFC algorithm. When adding small position errors to the SMF before running the algorithm, the achieved normalized intensity remains on the order of the performance of the perfectly aligned case. For instance, a position displacement of the order 1% of $\lambda/D$ causes the algorithm to converge slower if the offset is not accounted for in the model and the final SMF normalized intensity is within a factor of two compared to the perfectly aligned case.

The response to tip-tilt errors after the null is produced is significantly worse. We simulate this by generating the null at a nominal position and, with the DM solution applied, introduce small displacements to the SMF without further wavefront correction. Figure 5 shows eight cases corresponding to different displacement directions. We find that the response in the terms of normalized intensity is very chromatic under tip-tilt errors and that the direction of displacement has a significant effect on the degradation of the normalized intensity (compare, e.g., 0 deg and 90 deg displacements). This is due to the structure of the phase that the DM solution induces in the image plane; some directions will still have a phase pattern that nulls the overlap integral. However, most directions of displacement are very sensitive, with a deterioration of one order of magnitude in the normalized intensity for a displacement of 0.2% of $\lambda/D$, and over two orders of magnitude for a displacement error of 1% of $\lambda/D$.

The phase solution that the DM induces at the tip of the fiber has to be asymmetric (see Fig. 4), as discussed by Por and Haffert,[16] this asymmetry can be of first-order, second-order, etc. depending on the phase structure that achieves the null. In general, the phase structure found by the algorithm is of first order, i.e., a phase ramp across the fiber tip, which causes a significant leak of light for small misalignments, and thus it is a more sensitive solution to tip-tilt errors. Although this is a limitation with respect to conventional EFC, future work will explore methods to reduce the tip-tilt sensitivity of the fiber-based solutions, including using a controller that initially reduces the intensity at the fiber before finding the best null using the overlap integral. For the purpose of this work, we demonstrate the current algorithm without attempting to reduce the sensitivity of the solution to tip-tilt errors.

## 6 Laboratory Setup

To validate the algorithm we performed experiments using the HCST-T,[9] an optical testbed consisting of an AO system, a coronagraph, and an FIU (see Fig. 6). The optics are mainly off-the-shelf transmissive lenses that are readily available.

For the monochromatic tests, we use a laser diode at 635 nm; for the broadband tests we use a supercontinuum white light laser source (NKT Photonics SuperK EXTREME), filtered to provide a $\Delta\lambda/\lambda = 8\%$ bandpass at 625 nm. The light is fed into HCST-T by two source fibers (Thorlabs SM600 fibers), which simulate the star and planet. For this work, we will only make use of the star source. A telescope simulator, with an aperture diameter of 4 mm, images the simulated star. In the AO system, a DM (Boston Micromachines Corporation multi-DM) controls the incoming wavefront. The DM has a continuous membrane surface with $12 \times 12$ actuators and a 400-$\mu$m actuator pitch. The beam illuminates a circular region 10 actuators in diameter. The specified average step size of the actuators is <1 nm.

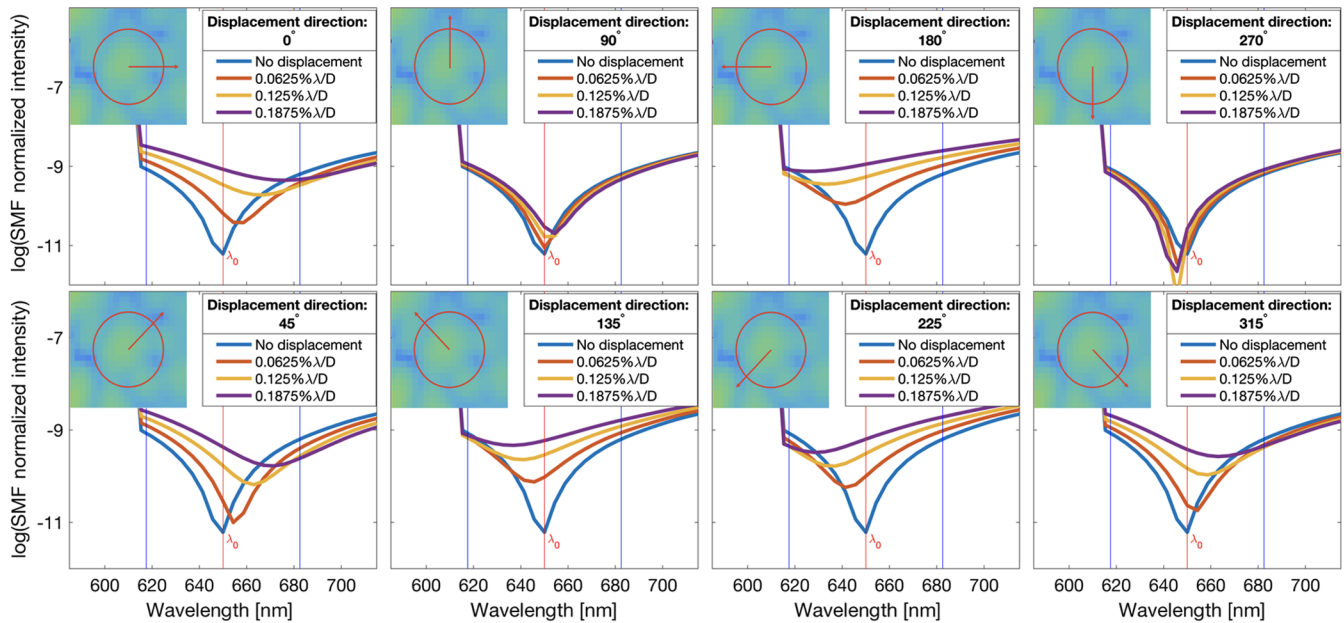

**Fig. 5** Simulation showing the effect of moving the SMF from its original position where the starlight is nulled in eight directions. We find the change in normalized intensity is chromatic and direction dependent. The red and blue lines indicate the central wavelength and the limits of the controlled bandwidth, respectively.

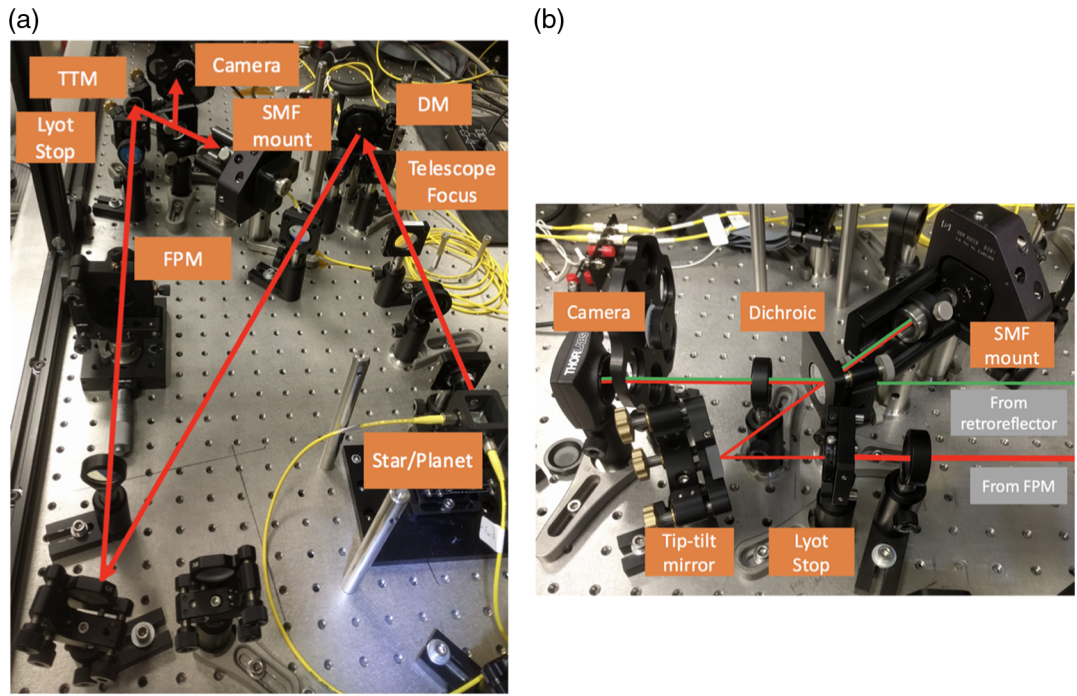

**Fig. 6** (a) The HCST-T layout consists of two fiber-coupled sources to simulate a star and planet, an AO system with a DM, a coronagraph with an FPM and a Lyot Stop, and an FIU with a TTM, SMF mount, and a tracking camera. (b) At the FIU, the beam is steered by the TTM to align it with the SMF while the dichroic sends part of the incoming light to the tracking camera. The SMF can be used to back-propagate light into the system where it is reflected by the dichroic to a retroreflector such that the SMF is also imaged by the tracking camera for alignment and calibration purposes.

The beam then passes through a three-plane coronagraph, where the light is focused on to the focal plane mask (FPM). Our setup is equipped with a vortex coronagraph, which enables high-throughput, high-contrast imaging at small angular separations.[17,18] We use a charge 4 liquid crystal polymer vector vortex mask, which applies a phase ramp at the focus of the form $e^{\pm i4\theta}$. This FPM is optimized around 600 nm. The quality of a vector vortex phase mask is characterized by measuring the transmission between parallel circular polarizers to estimate the fraction of starlight with the incorrect phase. For the mask used here, the leakage is less than 0.15% and 0.11% for 635 and 625 nm, corresponding to the central wavelengths of the monochromatic and broadband experiments, respectively. The beam is then collimated and clipped by an adjustable iris that serves as the Lyot stop and blocks ~15% to 20% of the full pupil area. The beam magnification between the DM and Lyot stop is 1:1.

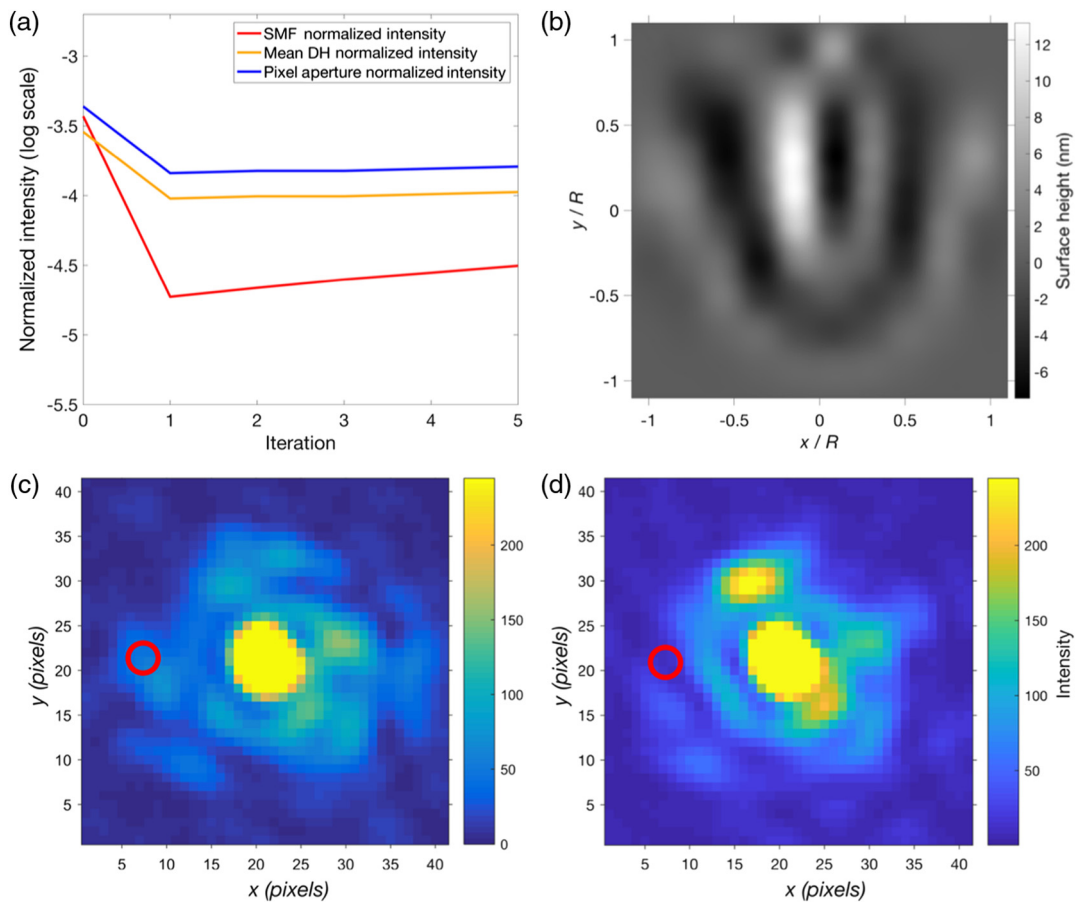

**Fig. 7** Laboratory results for conventional EFC experiments on HCST-T. (a) normalized intensity versus iteration, (b) the DM shape solution according to our model, and (c)-(d) the coronagraphic PSF before and after correction, respectively. The normalized intensity achieved is likely limited by the high levels of aberration and the model uncertainty. In (a), the SMF normalized intensity is also plotted although the control is entirely done with the camera. The SMF normalized intensity is always better thanks to the modal selectivity of the SMF. The final solution of the DM (b) has the expected shape, given the small size of the DH, with a distinct sinusoidal shape at the spatial frequency corresponding to the position of the DH. In (c) and (d), a 3 × 3 pixel box located at the center of the red circle is the control area, or the DH in which EFC is trying to null.

Finally, a tip-tilt mirror (TTM) sends the beam into the FIU. The FIU system is nearly identical to the one described by Mawet et al.[9] (see Fig. 6). The TTM is actuated in order to accurately align the beam to the SMF. A dichroic lets the majority of the light go through to the SMF and reflects some light to the tracking camera (Thorlabs CMOS DCC1545M). The camera is used for positioning the fiber, aligning the coronagraph mask, and to perform conventional pixel-based EFC, as shown in Sec. 7.1. The SMF is mounted on a five-axis stage (Newport 9091) behind a 7.5-mm focal length lens. At the output end of the SMF, the power coupled into the SMF is measured with a silicon photodiode (FEMTO OE-200-SI).

We measured the throughput of the FIU to be 55% by comparing the power measured upstream of the focusing lens to the output of the SMF. However, the ideal coupling for a circular aperture into an SMF with perfect optics is 82%. We identify various sources to account for the loss of throughput: the transmission of the focusing lens, transmission losses in the SMF, and the mismatch between the focal ratio, $F\#$, of the incoming beam and the optimum for our SMF. To isolate these effects, we removed all of the optics between the source and FIU and measured the low order aberrations upstream of the focusing lens using a Shack–Hartmann wavefront sensor (Thorlabs WFS150-5C). In our numerical simulation, we introduced the measured aberrations to the simulated wavefront and took into account the mismatch between the $F\#$ of the last lens and the optimal $F\#$ for the fiber. The result was consistent with the measured losses in our system; the main cause of throughput loss being the coupling of the suboptimal $F\#$.

## 7 Results

### 7.1 Conventional EFC Tests

In order to assess the wavefront control capabilities of HCST-T, we first performed conventional camera-based EFC tests. Although speckle nulling has been previously demonstrated by our team using this setup, both using the camera and an SMF,[9] EFC is a significantly different algorithm. EFC relies on an estimate of the electric field at the image plane, an accurate model of the system, and low level of aberrations so that the response of the system to changes in the plane of the DM is linear in image plane field amplitude.[11] Given that HCST-T consists of off-the-shelf transmissive optics, the low order aberration regime is not guaranteed; indeed, our starting focal plane location at $4\lambda/D$ is of the order of $10^{-4}$ normalized intensity. Besides, although the DM has a total of $12 \times 12$ actuators, the pupil is clipped at the DM plane and only $10 \times 10$ actuators are available, therefore, the control radius is limited to $5\lambda/D$.

In Fig. 7, we show the results for the tests on conventional camera-based EFC. The control area, or DH, is a $3 \times 3$ pixel box centered at approximately $4\lambda/D$ from the PSF; the resolution at the camera is of 3.2 pixels per $\lambda/D$ approximately. We achieve a modest normalized intensity of $10^{-4}$. The limited performance is attributed to having a low fidelity model of the physical system, to both estimate the electric field and to compute the G matrix of the system. The shape of the DM, according to our model, agrees well with our expectation for a small DH at $4\lambda/D$, consisting of a distinct sinusoidal shape at the corresponding spatial frequency [see Fig. 7(b)]. The discrepancy between this shape and the shapes found via simulations [see Fig. 2(c)] can be explained by the fact that in the laboratory, after achieving a certain normalized intensity, an order of magnitude lower than the vicinity of the control area, the electric field becomes increasingly hard to sense: the intensity modulation starts to worsen, and the limited dynamic range of the detector makes it harder to calibrate the probes. Therefore, the algorithm fails to find a better shape for the DM. The RMS surface height of the DM solution within the pupil is 3.3 nm RMS.

### 7.2 Monochromatic Light Results

Figure 8 shows a laboratory demonstration of the new algorithm, in which a suppression of a factor of ~100 is achieved through the SMF. The final SMF normalized intensity is $3 \times 10^{-6}$. Figures 8(c) and 8(d) show the coronagraphic PSF before and after correction, and Fig. 8(b) shows the solution for the DM. The SMF is placed at approximately $4\lambda/D$ to avoid PSF distortion effects at smaller angular separations due to the FPM and to stay within the $5\lambda/D$ control radius afforded by the available $10 \times 10$ actuators at the DM plane.

The improved performance of the new algorithm compared to the tests presented in Sec. 7.1 can be explained by the two reasons discussed in Sec. 5. Indeed, the nature of the problem is different and the DM is only restricted to one control element. Furthermore, the sensing of the electric field is more favorable when using the new algorithm. This is because the probes used to sense the overlap integral are just sinusoids on the DM, or satellite speckles at the image plane. Their effect on the change on the overlap integral is more robust, given the modal selectivity of the SMF. As found via simulations (see Sec. 5), the values of the actuator strokes on the DM are significantly smaller, with respect to the solution for conventional camera-based EFC. The RMS surface height of the DM solution within the pupil is 2.2 nm RMS. Hence, the effect of this DM solution on the Strehl ratio will be more favorable with respect to conventional EFC. However, the presence of the vortex coronagraph, and the fact that we work at $4\lambda/D$, will degrade the coupling efficiency into the fiber. The difference between achieved normalized intensity for the intensity and for the SMF normalized intensity as expected from the simulations is reproduced in the laboratory [see Fig. 8(a)].

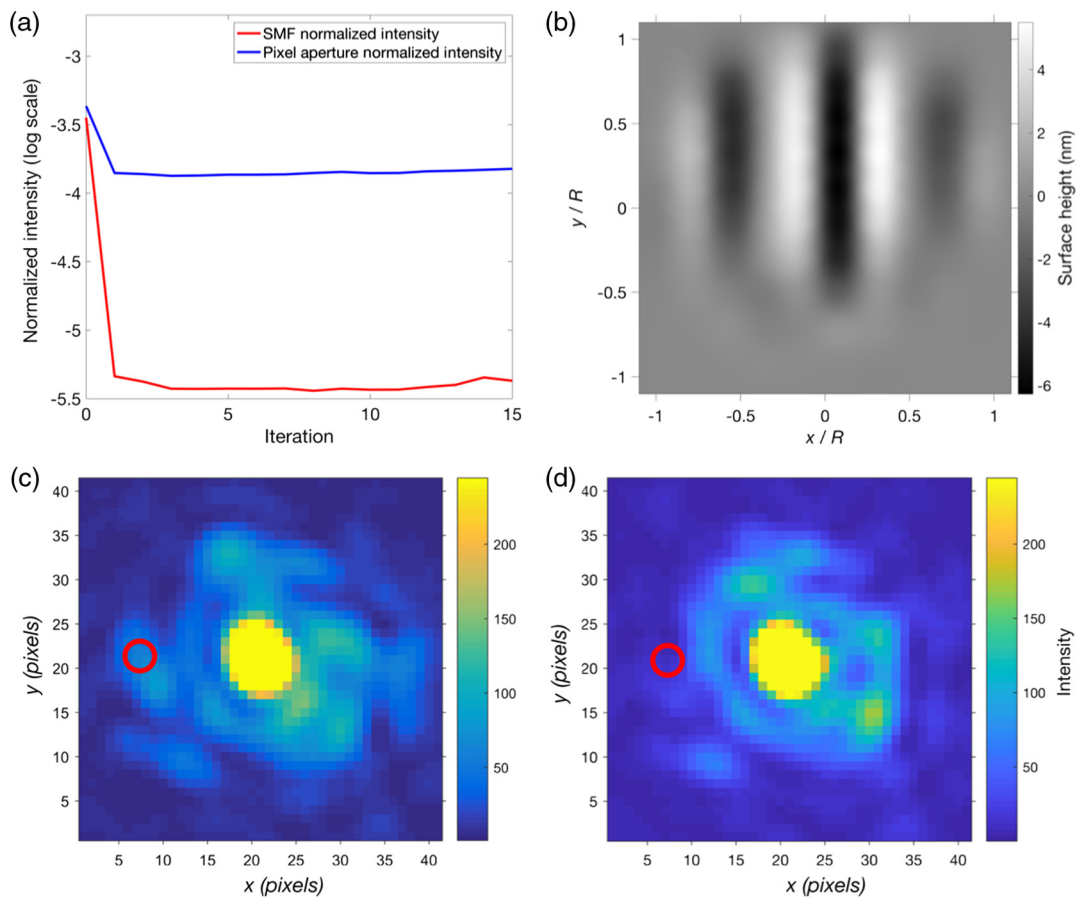

**Fig. 8** Same as Fig. 7, but for monochromatic fiber-based EFC experiments. The main features of the curves in (a) are as predicted by the simulations: the algorithm reaches its final normalized intensity after only a few iterations and the normalized intensity on the camera is >10× the SMF normalized intensity. The DM solution in (b) is very similar to the solutions found via simulations. The main features of the DM shape are found at the first iteration, the following iterations are just minimal adjustments. The red circle in (c) and (d) indicates the position of the SMF.

### 7.3 Polychromatic Light Results

We performed polychromatic light experiments with the new algorithm using a $\Delta\lambda/\lambda = 8\%$ bandwidth centered at $\lambda_0 = 625$ nm. The algorithm remains unchanged; i.e., it only aims at controlling the central wavelength while the full band of the light is fed into system at once. The only change in the setup is the use of a different light source; we connected the supercontinuum source with a 50-nm bandpass filter.

We show in Fig. 9 that we obtained an SMF normalized intensity of $1.6 \times 10^{-5}$, a degradation of a factor of 5 in terms of normalized intensity with respect to the monochromatic case. Due to the larger bandwidth, the effect of the FPM is significantly limited, and more light passes through the mask due to the chromatic leakage. However, the algorithm is still able to control some of the light as can be seen from the contrast curves. The RMS surface height of the DM solution within the pupil is 2.0 nm RMS.

### 7.4 Considerations on the Control Performance

The achieved normalized intensity in the experiments presented in the previous sections is far from reaching the noise floor of the detector, which sets the limit of SMF normalized intensity to a level of $\sim 1 \times 10^{-12}$. Furthermore, the simulations for the new algorithm predict an almost perfect suppression of residual starlight in monochromatic light. The limitations on the performance of our experiments can be explained by discrepancies between the model used in the algorithm and the real optical system. EFC needs an accurate model to build the $G$ matrix of the system and to get an accurate estimation of the electric field (or overlap integral). Discrepancies between model and optical system are a commonly known problem when implementing EFC, and an important limitation for the achievable contrast.[19]

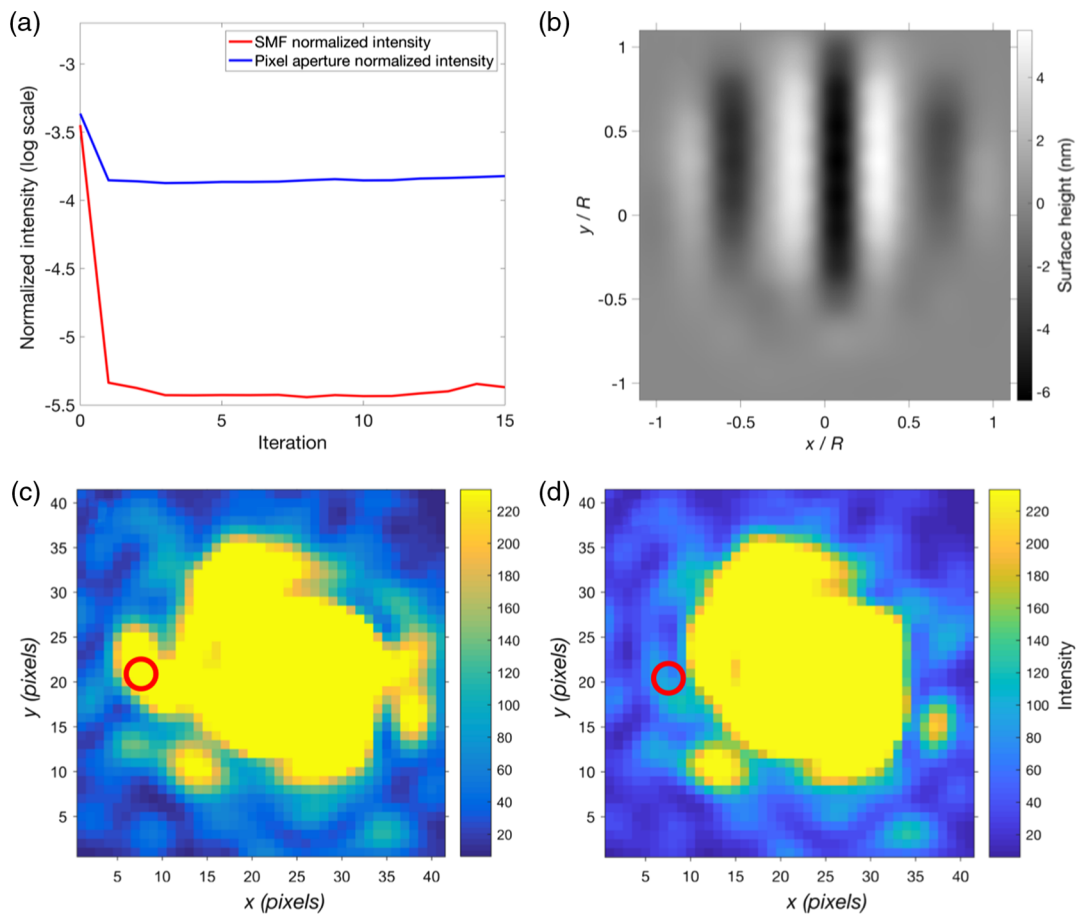

**Fig. 9** Same as Fig. 8, but for polychromatic light. The normalized intensity achieved is 5 times worse than the monochromatic experiment. The fact that we are controlling only the central wavelength is a primary cause of the deteriorated performance. The DM solution in (b) remains almost identical to the case of monochromatic light. The camera images of the coronagraphic PSF (c) and (d) show how more light gets leaked into the image plane.

We identify some specific sources of model uncertainty on HCST-T:

- The vector vortex coronagraph imparts conjugated $e^{\pm i4\theta}$ phase ramps on input orthogonal circular polarization states. Our EFC-based controller can only control one state and thus one phase ramp at a time, leaving noncommon path aberrations uncorrected.

- The quality of the transmissive optics coupled with the uncertainties on the alignment of the system. Since the model used for running the control algorithm relies on Fourier transforms for a flat wavefront, the model uncertainties arising from the aberrations on the optics and the errors in the alignment are not properly accounted for.

- The Lyot stop position and shape. There is uncertainty on the exact position of the Lyot stop with respect to the conjugated entrance pupil plane. In our model, the aperture plane is perfectly conjugated with the Lyot stop plane. Furthermore, the aperture of the Lyot stop is a manual, adjustable iris, which results in an uncertainty on the amount of light clipped by the Lyot stop. In the model, the beam is assumed to be perfectly circular, but in the real setup the beam may be somewhat noncircular. This has an effect on the shape of the PSF and, thus, on the coupling with the SMF.

- The DM shape uncertainty. Although we do not find any limitation in our simulations in monochromatic light, which assumes a perfect DM, there may be shape errors in the form of actuator response errors, or interactuator coupling related errors. The influence function used is not measured from our DM; rather, we use a smooth shape similar to a Gaussian.

- The uncertainties regarding the coupling of the light with the SMF. Although we can account for the losses in the coupling at the FIU, as discussed in Sec. 6, the effects on the algorithm of factors such as the modeling of the fundamental mode of the SMF, or the photonic effects in the SMF itself, are poorly understood.

Some other suspected reasons for the limitation in the laboratory performance of the algorithm are:

- The DM control authority. A $12 \times 12$ actuator DM is severally limited in the range of shapes it can reproduce, especially at high spatial frequencies near the Nyquist limit.

- The limitations on the SMF position. As discussed in Sec. 7.2, the range of positions in which we can place the SMF with respect to the PSF is limited by the number of DM actuators across the pupil and the inner working angle of the FPM. In practice, we place the SMF at approximately $4\lambda/D$ from the central PSF. At this spatial frequency, if we apply a satellite speckle with the DM in both the laboratory and the model, we can see a significant difference in the shape of the speckle due to the effect of the FPM.

- The stability of the setup. Although HCST-T is equipped with a full solid enclosure and the PSF in the camera appears to be very stable, the coupling into the SMF is extremely sensitive. The deviations of the SMF from its original position are not monitored, but the effect of changes in the SMF position could be detrimental, especially in the sensing stage. In Sec. 5.1, we found the null through the SMF to be very sensitive to jitter, which imposes strict requirements on postcoronagraph tip-tilt control.

## 8 Perspectives

After having demonstrated this new algorithm on the HCST-T, we plan to move the experiment to the superior HCST-R.[14] Equipped with custom reflective optics and a BMC kilo-DM with 34 actuators across the pupil, HCST-R has excellent potential for exploring high-contrast technologies. We have achieved a normalized intensity of $5 \times 10^{-8}$ using a simple camera-based speckle nulling technique, our plan is to include an FIU at the image plane of HCST-R and achieve very high contrast in polychromatic light through an SMF.

A Kalman filter was implemented for speckle nulling by Xin et al.,[20] in which the control history, and previous measurements, were used to achieve a more stable null through the SMF and an overall better normalized intensity. A Kalman filter estimator for EFC was demonstrated by Groff and Kasdin,[21] for a faster suppression of the electric field of the starlight in an EFC dug DH, with further improvement by adding an extended Kalman filter by Riggs et al.[22] This technique may be directly applied to the case of EFC for an FIU, and we plan to demonstrate this on HCST-R. Predictive control is particularly important in the presence of atmospheric turbulence and other types of disturbances such as vibrations and thermal drifts; a Kalman filter approach, which can account for the nature of the speckle evolution in the image plane, is a very promising technique.

We also plan on demonstrating this technique on sky with the Keck Planet Imager and Characterizer (KPIC)[23] at the W.M. Keck Observatory. KPIC consists of a series of upgrades to the Keck II AO system and instrument suite, including an FIU to high-resolution infrared spectrograph NIRSPEC. In addition to its unique science capabilities, KPIC is also intended as a path to mature key technologies, such as high dispersion coronagraphy (HDC),[9,24–28] for future space-based telescopes and large ground-based telescopes such as the Thirty Meter Telescope. KPIC is a perfect instrument to test this algorithm on sky.

In the limiting case where stellar photon noise originating from quasistatic aberrations is dominating (e.g., HR8799's planet infrared spectroscopy with KPIC), the corresponding exposure time gain is $\tau \propto \eta_s/\eta_p^2$ (see Ref. 29), where $\eta_s$ and $\eta_p$ are the fraction of residual star and detected planet light, respectively. The achieved stellar signal suppression of $\sim 100$ shown in this paper would translate into a reduction of $\sim 100$ in necessary exposure time. This algorithm, if running fast enough, and/or combined with a Kalman filter could also address dynamic atmospheric residuals. This will be the subject of a forthcoming paper.

## 9 Conclusion

We have presented an algorithm, based on EFC, to achieve improved suppression through a SMF. We performed simulations to assess the performance of the algorithm and its sensitivity to position errors and jitter, and tested it in the laboratory, where we obtained a normalized intensity through the SMF of $3 \times 10^{-6}$ in monochromatic light at 635 nm, and $2 \times 10^{-5}$ in 8% broadband light at 625 nm. The wavefront control algorithm presented here is designed to take advantage of the SMF's spatial selectivity, thus is perfectly suited for an HDC system.[9,28] The promising results obtained from simulations, and the lessons learned from applying EFC in the laboratory on HCST-T, will help us achieve the significantly deep contrast levels on our improved HCST-R testbed. The stellar suppression gains obtained by this technique directly reduce the exposure time needed for stellar photon noise limited cases (see Sec. 8), since the Strehl ratio is practically unaffected by the DM solution (see Sec. 7.2). Applying this algorithm in practice on future telescopes may enable the detection of spectral signatures associated with individual molecules and potential signs of life.[3]

## Acknowledgments

This research was partially supported by the National Science Foundation AST Grant 1710210.

Biographies of the authors are not available.